**Multi-Agent Based Simulation for Decentralized Electric Vehicle Charging Strategies and their Impacts**

Christensen, Kristoffer; Jørgensen, Bo Nørregaard; Ma, Zheng Grace

Go to publication entry in University of Southern Denmark's Research Portal

# Multi-Agent Based Simulation for Decentralized Electric Vehicle Charging Strategies and their Impacts

Anonymized

**Abstract.** The growing shift towards a Smart Grid involves integrating numerous new digital energy solutions into the energy ecosystems to address problems arising from the transition to carbon neutrality, particularly in linking the electricity and transportation sectors. Yet, this shift brings challenges due to mass electric vehicle adoption and the lack of methods to adequately assess various EV charging algorithms and their ecosystem impacts. This paper introduces a multi-agent based simulation model, validated through a case study of a Danish radial distribution network serving 126 households. The study reveals that traditional charging leads to grid overload by 2031 at 67% EV penetration, while decentralized strategies like Real-Time Pricing could cause overloads as early as 2028. The developed multi-agent based simulation demonstrates its ability to offer detailed, hourly analysis of future load profiles in distribution grids, and therefore, can be applied to other prospective scenarios in similar energy systems.

## 1    Introduction

The transformation of the electricity grid into a Smart Grid integrates digital energy solutions designed to address the challenges posed by the shift towards carbon neutrality. This transition, aligned with the European Union's goal for a carbon-neutral society by 2050 [1], necessitates the electrification of the transportation sector. The rapid replacement of conventional vehicles with Electric Vehicles (EVs) is critical but demands an enhanced charging infrastructure to meet the increased electricity requirements [2]. Yet, existing grids are currently ill-equipped for this surge, presenting significant challenges.

Digital solutions that leverage EV charging flexibility, such as smart EV charging systems and intelligent chargers, can significantly reduce the costs associated with integrating EVs into the electrical grids [3, 4]. However, the rising number of EVs risks causing substantial grid overloads under existing conditions. Distribution System Operators (DSOs) must choose between costly grid upgrades or adopting dynamic tariff schemes that utilize energy flexibility to manage and potentially delay these upgrades.

The literature and interviews with DSOs in Denmark highlight three main challenges associated with the rise of EVs and their charging in distribution grids: the unclear impacts of implementing Dynamic Distribution Tariffs (DDTs) with increasing numbers of EVs [5, 6], and the unpredictable consequences of adopting EVs, DDTs, and various charging strategies. These strategies aim to harness the flexibility of EVs, yet the comprehensive effects on all ecosystem stakeholders—including DSOs, EV users, and service providers—are not fully understood, complicating the strategic decisions needed to prevent economic losses and avoid unnecessary expenditures.

Additionally, the adoption of dynamic electricity pricing structures, such as Time-of-Use tariffs and hourly electricity prices, is influencing end-consumers' awareness and consumption patterns. These pricing models encourage consumers to be more conscious of their electricity usage times and volumes, thus adding complexity to EV charging, which transcends the simple act of plug-in and charging. Prior to 2021 in Denmark, many consumers employed a straightforward plug-in and charge approach, largely due to a lack of awareness of dynamic pricing [7]. This strategy, referred to as Traditional charging, is evolving post-2021 with the availability of hourly price settlements, potentially altering consumer behavior.

Furthermore, Smart charging, as defined by [8] and [9], involves the intelligent management of EV charging through connections between EVs, charging operators, and utility companies facilitated by data links. This paper categorizes charging strategies based on their control architecture, data requirements, and charger intelligence. Several EV charging algorithms are proposed in the literature, e.g., [10-13], however, according to a feasibility evaluation [14], the short-term feasible decentralized EV charging strategies identified include Time-of-Use (ToU) Pricing, Real-Time Pricing (RTP), and Timed charging, the latter often managed via smartphone apps but not considered "smart" in simpler implementations.

This study utilizes a multi-agent based simulation model to explore these dynamics in a case study of Strib, Denmark, with 126 households. The simulation aims to realistically represent the EV home charging ecosystem, informed by real-world data on EV adoption rates, vehicle models, driving patterns, and baseline consumption.

The remainder of this paper is structured as follows: The Methodology section presents the multi-agent simulation framework, detailing the parameters and configurations employed for modeling decentralized EV charging strategies. This is followed by the Case Study and Scenarios section, which applies these strategies within a specific context. Next, the Results and Discussion section evaluates these strategies in comparison to traditional charging methods. The paper concludes with the Conclusion section, summarizing key findings and implications.

## 2    Methodology

This paper begins by analyzing the components of the EV home charging ecosystem—actors, objects, roles, and interactions—using the business ecosystem architecture for EV home charging as outlined in [15]. It then details how these elements are integrated into a multi-agent based simulation, following the methodology described in [16]. The process of translating actors, roles, and interactions into agents, interfaces, and agent communications within the model architecture is elucidated. This simulation is developed using the multi-method simulation tool, AnyLogic.

## 2.1 Agent Types

The agents represent actors and objects in the ecosystem. The interaction in the business ecosystem is between roles. Therefore, the roles are represented by Java interfaces holding the interactions related to the role. The agents implement the respective interfaces corresponding to the actor/object and the associated roles.

The selected charging strategies (RTP and ToU Pricing) are implemented into the simulation model developed in the software AnyLogic. This section introduces and explains the content of all agent roles relevant to the controlled charging of the EV. That includes the agents and interfaces shown in Table 1. The decentralized charging strategy agents are located at the ChargingServiceProvider, whereas, in reality, decentralized logics/algorithms are located at the individual charging controls. This design is selected for modeling convenience and simulation performance, as locating them at the ChargingBox agents would result in many of the same agents performing the same actions. This design, instead, locates the logic in one place (ChargingServiceProvider agent) for all ChargingBox agents to call its functions to receive charging schedules.

**Table 1.** Relevant agents related to charging strategy implementation.

| Agent name | Roles | Implemented interfaces |
|---|---|---|
| DSO | Distribution system operator | • I_DSO |
| ChargingBox | Charging Control | • I_ChargingControl <br> • I_Helper_Message <br> • I_Appliance |
| ChargingBoxSupplier | Service Provider | • I_AppServiceProvider |
| ElectricVehicle | Electric Vehicle | • I_ElectricVehicle |
| DomesticConsumer | Electric Vehicle User | • I_ElectricityConsumer <br> • I_ElectricVehicleUser <br> • I_ElectricityLoad <br> • I_MeterOperator |
| RealTimePricing <br> ToUPricing | Decentralized charging strategy | • I_Strategy_RTP <br> • I_Strategy_ToU |
| ChargingServiceProvider | Charging service provider | • I_ChargingServiceProvider |

## 2.2 Agent Logic

**Real-Time Pricing.** The RTP strategy agent is named "RealTimePricing" in the model and is referred to as the RTP agent in this research. The RTP strategy algorithm's logic is shown in **Fig. 1**. The green square represents the logic located within the RTP agent. The blue square represents the algorithm's required input from the EV and/or EV user.

**Time-of-Use Pricing.** The ToU Pricing charging strategy agent is named "ToUPricing" in the model. The logic for this agent is identical to the RTP, except that instead of the day-ahead hourly electricity price, the ToU logic uses the tariff schedule from the electricity supplier (#5 in **Fig. 1**). When prices are sorted by the lowest price first, the second sorting condition is the time (e.g., if only one hour of charge is needed within the cheapest period, the schedule will contain the first hour in the period).

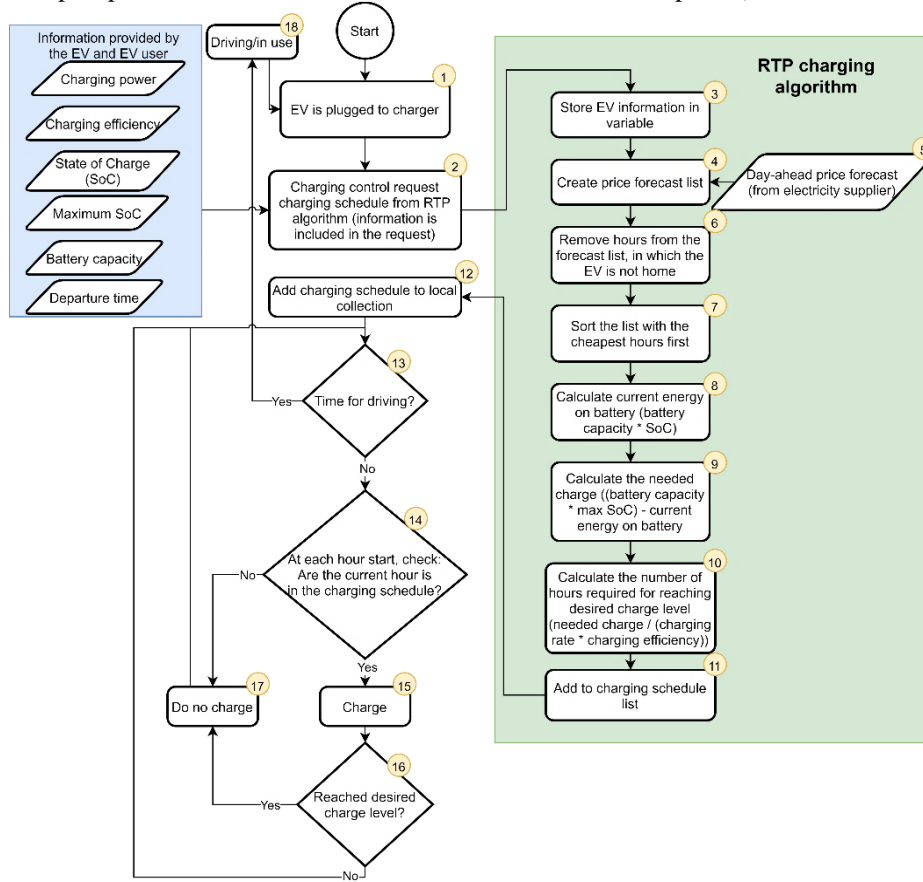

**Fig. 1.** The EV charging strategy agent logic.

The logic flow is illustrated in Fig. 1 as a flowchart:

(#1 in Fig. 1) The charging with the RTP algorithm starts by plugging the EV into the charging box.

(#2) The charging box agent sends a request for a charging schedule to the RTP agent.

(#3) The request includes the information seen in the blue box. The EV information is stored in local variables in the RTP agent.

(#4-5) The RTP algorithm uses day-ahead forecast prices for deciding charging times. The day-ahead prices are collected from the electricity supplier agent and stored in a local list.

(#6) The hours when the EV is not at home (known based on the departure time) are removed from the price forecast list.

(#7) The list with prices and corresponding times is sorted in ascending order, i.e., the lowest price first.

(#8) The current energy on the EV battery is calculated:

$$E_{bat} = Cap_{bat} \cdot SoC \tag{1}$$

(#9) Where $Cap_{bat}$ is the battery capacity, and $SoC$ is the current State-of-Charge (SoC). The energy required to reach the desired SoC is calculated:

$$E_{bat,req} = Cap_{bat} \cdot SoC_{max} - E_{bat} \tag{2}$$

(#10) $SoC_{max}$ is the maximum SoC for this EV user (desired SoC). From the required energy, the number of charging hours is calculated:

$$h_{req,charge} = \frac{E_{bat,req}}{P_{charge} \cdot \eta_{charge}} \tag{3}$$

Where $P_{charge}$ is the charging power and $\eta_{charge}$ is the charging efficiency (84%).

(#11) The number (corresponding to the required charging hours) is taken from the beginning of the sorted forecast list and stored in a separate list.

(#12) This list corresponds to the charging schedule and is sent to the charging box agent.

(#13-18) In the last part of the flowchart, the EV checks each hour if the time is in the charging schedule (charging if the time is in the schedule) until the desired SoC is reached and it is time for departure.

## 3 Case Study and Scenarios

### 3.1 Case Introduction

This study focuses on a distribution grid in Strib, Denmark, serving 126 residential consumers. The Danish DSO TREFOR, which operates in this area, provided the consumption data. Only data from 2019 is used, excluding the COVID-19 impacted years of 2020 and 2021. After cleaning the data to remove households with EVs, PVs, heat pumps, electric heating, and missing data points, the final dataset covers 97 residents. To maintain the study's scale at 126 households, additional data from 29 randomly selected residents in Nr. Bjert was added. This adjusted dataset, free from distributed energy resources like heat pumps, helps simulate typical consumer consumption patterns, detailed in [16].

### 3.2 Scenarios and Simulation Data Input

The paper evaluates the ecosystem impacts of decentralized EV charging strategies, particularly comparing Real-Time Pricing (RTP) and Time-of-Use (ToU) Pricing against traditional charging schemes. The scenario relies on an adoption curve and electricity prices drawn from historical data and hourly price schemes under the DDT—Tariff Model 3.0. Various experiments are designed to test the effects of RTP and ToU Pricing, outlined in Table 2.

Simulation inputs include EV models, CO2 emissions, and driving patterns (departure, arrival, and distance traveled). The electricity pricing data, based on hourly spot prices from the Nordpool market for 2015-2018, supports the simulation's hourly resolution. As of January 1, 2023, TREFOR and eight other DSOs transitioned to Tariff Model 3.0, which features four distinct tariff periods year-round, replacing the simpler winter-only high-tariff model [17]. The analysis in [18] identifies Time-of-Use Pricing as the most feasible DDT for the Danish system, informed by a comprehensive assessment of 16 unique tariffs considering technical, economic, social, and regulatory factors.

**Table 2.** Experiments for the various decentralized charging strategy scenarios.

| Experiment Name | Purpose | Parameter Setup |
|---|---|---|
| Real-Time Pricing with Tariff Model 3.0 | To examine the effects of Real-Time Pricing (RTP) under Tariff Model 3.0 (TM3). | RTP charging strategy, TM3 |
| Real-Time Pricing – Distance Optimization (Min SoC 20%) with TM3 | To compare the performance of RTP with Distance Optimization and a minimum SoC of 20% to Time-of-Use (ToU) Pricing under similar conditions. | RTP charging strategy, Distance optimization, Min SoC 20%, TM3 |
| Time-of-Use Pricing with TM3 | To assess the impact of the ToU Pricing strategy under TM3. | ToU Pricing strategy, TM3 |
| Time-of-Use Pricing 80% SoC | To test effects of limiting EV's SoC to 80% to reduce battery degradation. | ToU Pricing strategy, Max SoC 80% |
| Time-of-Use Pricing – Distance Optimization (Min SoC 20%) | To evaluate the effects of ToU Pricing with Distance Optimization and a minimum SoC of 20% for battery management or reserve capacity. | ToU Pricing strategy, Distance optimization, Min SoC 20% |

## 4        Results and discussion

The following sections show the key results and relevant figures for each distinct strategy and its associated experiments. The experiments are compared to the Traditional charging strategy with their respective electricity price structure setup. For instance, strategies employing an hourly electricity price scheme and TM3 are compared with the "Tariff Model 3.0 and Hourly Electricity Price Scheme".

### 4.1        Real-Time Pricing Scenario results

The Real-Time Pricing (RTP) charging strategy, by default, charges EVs during the cheapest hours between arrival and departure. Table 3 summarizes the key results from the experiments comparing the RTP strategy with Tariff Model 3.0 (TM3) to traditional charging methods.

**Table 3.** Key results for Real-Time Pricing charging strategy experiments with Tariff model 3.0 and comparison with the Traditional charging strategy.

| KPIs | Traditional Charging | RTP with TM3 | RTP - Dist. Opt. (Min 20% SoC) | % Diff (RTP vs Trad.) | % Diff (RTP - Dist. Opt. vs Trad.) |
|---|---|---|---|---|---|
| Date of First Overload | 21 Oct. 2031, 4 PM | 30 Mar. 2029, 4 AM | 3 Nov. 2029, 12 AM | -21.70% | -16.64% |
| Overloads in Following Year | 60 | 248 | 79 | +313.33% | +31.67% |
| Avg. Charging Cost (DKK/kWh) | 1.3724 | 1.1871 | 1.1855 | -13.50% | -13.62% |
| Avg. Electricity Bill (DKK) | 11,178.06 | 10,251.66 | 10,231.31 | -8.29% | -8.47% |
| Avg. CO2 Emissions (kg) | 53.4965 | 46.4479 | 46.4617 | -13.18% | -13.15% |
| EV User Dissatisfaction | 63 | 80 | 56 | +26.98% | -11.11% |
| Load Factor | 0.2048 | 0.1002 | 0.1315 | -51.07% | -35.79% |
| Coincidence Factor | 0.4274 | 0.8063 | 0.8072 | +88.65% | +88.86% |
| DSO Revenue (DKK) | 133,758.44 | 71,347.63 | 71,166.34 | -46.66% | -46.79% |
| Total EVs at First Overload | 85 | 39 | 48 | -54.12% | -43.53% |

Note: Coincidence factor is calculated for the day with the first overload.

The results show that the implementation of TM3 would increase the incentives for the users to adopt the RTP strategy and decrease the DSO's tariff revenue. The EV users' costs can be reduced with RTP and the TM3 compared to without TM3. At the same time, the tariff revenue is significantly reduced with an increase in overloads. The results show that the RTP with TM3 is not good for the DSO to avoid grid overloads.

Fig. 2 shows the system details for the day with the first overload. The figure shows that the total electricity price during the night becomes the cheapest due to the TM3. Meanwhile, the RTP algorithm selects the cheapest hours.

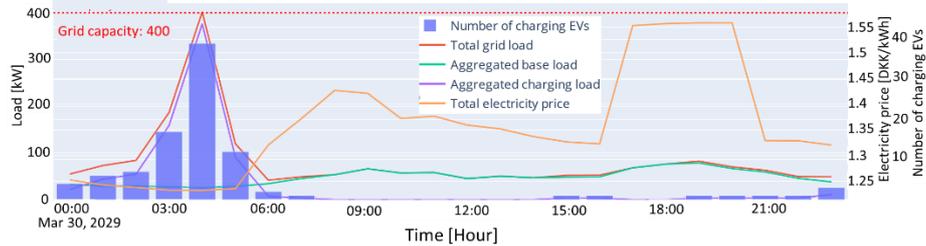

**Fig. 2.** Visualization of details on the day with the first grid overload (with the Real-Time Pricing charging strategy and Tariff Model 3.0).

The experiment of Real-Time Pricing – Distance optimization with minimum State-of-Charge of 20%" explores the effects of charging EVs according to the distance of the upcoming trip while maintaining a minimum SoC of 20%. Regardless of the short distances planned, the EVs are charged to achieve an SoC that is 20% higher than needed for the planned trip, ensuring a minimum SoC of 20% at all times. This approach, similar to maintaining a maximum SoC of 80%, aims to prolong battery life and prepare for emergency needs.

The results indicate that using the Real-Time Pricing (RTP) algorithm with this strategy lowers the total electricity bill but slightly increases the average charging cost compared to the standard RTP approach. Notably, the first grid overload is delayed by about 11 months, and there are five fewer overloads in the following year compared to the standard RTP. Additionally, this strategy allows the grid to manage 12 more EVs before reaching overload.

This method also decreases EV user dissatisfaction by 24 events compared to the standard RTP. However, despite these improvements, users of certain models like the Nissan Leaf continue to experience dissatisfaction with the RTP strategy when distance optimization is employed, mainly due to the insufficiency of the SoC for unexpected longer trips the following day.

### 4.2    Time-of-Use Pricing scenario results

The Time-of-Use (ToU) Pricing strategy, tested with configurations including an 80% maximum SoC and a distance optimization with a 20% minimum SoC, demonstrates financial performance comparable to Real-Time Pricing (RTP) but with quicker yet less frequent overloads. Table 4 provides a comparison of these results against traditional charging methods.

The ToU strategy results in earlier overloads in 2029 compared to 2031 for traditional charging, but experiences fewer subsequent overloads when employing distance optimization. Financially, all ToU experiments reduce the average charging cost by about 12.6% and electricity bills by up to 11.59%, while significantly reducing $CO_2$ emissions. The distance optimization strategy with a 20% minimum SoC shows the best performance in terms of delaying the first overload, handling more EVs on the grid, and improving load factors. While overall dissatisfaction increases under ToU strategies, it decreases significantly by 61.90% when using the 80% SoC strategy.

**Table 4.** Key results for Time-of-Use Pricing charging strategy with Tariff model 3.0 and comparison with the Traditional charging strategy.

| KPIs | Traditional Charging | ToU with TM3 | ToU 80% SoC | ToU - Dist. Opt. (Min 20% SoC) | % Diff (ToU vs Trad.) | % Diff (ToU 80% SoC vs Trad.) | % Diff (ToU - Dist. Opt. vs Trad.) |
|---|---|---|---|---|---|---|---|
| Date of First Overload | 21 Oct. 2031, 4 PM | 1 Jan. 2029, 12 AM | 1 Jan. 2029, 12 AM | 3 Nov. 2029, 12 AM | -23.74% | -23.74% | -16.64% |
| Overloads in Following Year | 60 | 216 | 215 | 102 | +260.00 % | +258.33 % | +70.00 % |
| Avg. Charging Cost (DKK/kWh) | 1.3724 | 1.1997 | 1.1995 | 1.1999 | -12.58% | -12.60% | -12.57% |
| Avg. Electricity Bill (DKK) | 11,178.06 | 10,302.03 | 9,882.5 | 10,304.08 | -7.84% | -11.59% | -7.82% |
| Avg. CO2 Emissions (kg) | 53.4965 | 45.8112 | 42.5423 | 45.6785 | -14.37% | -20.48% | -14.61% |
| EV User Dissatisfaction | 63 | 80 | 24 | 56 | +26.98 % | -61.90% | -11.11% |
| Load Factor | 0.2048 | 0.0997 | 0.0963 | 0.1305 | -51.32% | -52.98% | -36.28% |
| Coincidence Factor | 0.4274 | 0.8041 | 0.8041 | 0.8072 | +88.14 % | +88.14 % | +88.86 % |
| DSO Revenue (DKK) | 133,758.44 | 71,126.97 | 69,492.93 | 70,799.15 | -46.82% | -48.05% | -47.07% |
| Total EVs at First Overload | 85 | 37 | 37 | 48 | -56.47% | -56.47% | -43.53% |

Note: The coincidence factor is calculated for the day with the first overload.

The ToU Pricing charging strategy's logic starts charging at the beginning of the cheapest ToU period (in the period of which the EV is connected to the charger). If the ToU Pricing has one long cheap period, for instance, during the night, then the ToU Pricing strategy's logic will start charging the EV in the first hour of this period. The ToU Pricing strategy does not consider the raw electricity price as the logic is designed for an unkown electricity price scheme but with ToU tariffs.

For both summer and winter periods in the TM3, the cheapest period starts at midnight. Fig. 3 shows how all the EVs are charging from midnight, following the logic. Notice in Fig. 3 that there is one EV charging at 9 PM. This is because there are not enough hours in the cheapest period to satisfy its charging needs. Therefore, the EV charges the first hour of the second cheapest period, which would be 9 PM.

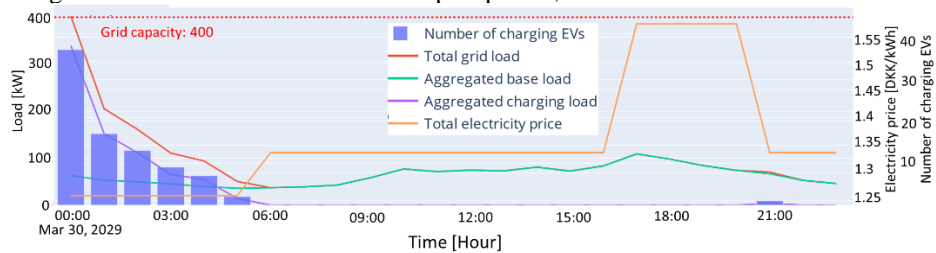

**Fig. 3.** Visualization of details on the day with the first grid overload (with the Time-of-Use Pricing charging strategy and Tariff Model 3.0).

For the ToU Pricing, as well as the RTP, EV users experience 17 events of more dissatisfaction compared to the baseline scenario. All dissatisfactions are only by Nissan

Leaf users. The RTP and ToU Pricing experience more dissatisfaction for Nissan leaf EV users because the algorithms for RTP and ToU Pricing calculate the charging schedule for the available hours. The algorithms currently use whole hours, meaning that the periods (from the EV arrives to the first full hour) are not used for charging, and the hours the EV departs are also not used in the charging schedule. If the EV's SoC is not satisfied after the last hour in the schedule, the EV will continue charging (without considering the cost) until it is satisfied or the EV departs. Therefore, this induces additional dissatisfaction for the Nissan leaf users as the extra charging within the arrival hour for the Traditional charging is often sufficient to satisfy the maximum SoC.

The realization of the algorithms in the research is based on the interpretation of the algorithm logic's description in the literature. In reality, the implementation of the algorithms should consider all available time. Therefore, further research should consider a better logic to improve the algorithms.

## 5      Conclusion

The increased adoption of Electric Vehicles (EVs) presents substantial challenges for Distribution System Operators (DSOs) in maintaining high security of supply without significant investments in grid reinforcement. This research demonstrates the potential of activating consumer flexibility through tariff schemes as a cost-efficient strategy for DSOs to manage increased demand.

This research applies decentralized EV charging strategies within a multi-agent simulation of an EV home charging ecosystem to explore the impacts from 2020 to 2032 at an hourly resolution. The results indicate that current electricity distribution grids are not equipped to handle the rising load from EVs without control measures. Traditional charging strategies lead to grid overloads by 2031 with 67% EV penetration, while the adoption of intelligent charging strategies like Real-Time Pricing (RTP) could precipitate these overloads as early as 2028.

This research addresses the research gap regarding the aggregated consequences of adopting EVs, Dynamic Distribution Tariffs (DDTs), and various charging strategies. The multi-agent based simulation developed in this research proves valuable for exploring other future and hypothetical scenarios in similar energy ecosystems.

Practically, the simulation model offers significant insights for stakeholders within the energy ecosystem. For instance, DSOs can utilize the model to assess load profiles on transformers and understand tariff revenue implications under different EV adoption curves and charging strategies. Specifically, the results suggest that Danish DSOs should reconsider the deployment of Tariff Model 3.0 if a high adoption of the RTP charging strategy is observed to better align their operational strategies with the evolving grid demands. For EV users, this model facilitates the evaluation of different EV charging programs offered by service providers, enhancing their decision-making process in selecting optimal charging strategies. Policymakers can leverage the findings to assess the impact of new regulations on the energy ecosystem and explore adjustments to tariff designs that could foster more effective and sustainable energy distribution systems.

Future research could consider investigating the effects of fluctuating electricity prices, particularly on industrial energy consumers. Additionally, as Tariff Model 3.0 is anticipated to reduce tariff revenue and increase overload frequency, further studies could explore the implications of centralized charging strategies and the potential need for grid extensions prior to 2030. This will aid in preparing for a transition that accommodates the rapid increase in EV adoption without compromising grid stability.

**Acknowledgments.** Anonymized

**Disclosure of Interests.** The authors have no competing interests to declare.

# References

1. European Comission. "Climate strategies and targets." https://ec.europa.eu/clima/eu-action/climate-strategies-targets_en (accessed July 21, 2022).
2. DTU and Dansk Elbil Alliance, "Sådan skaber Danmark grøn infrastruktur til én million elbiler," DTU Orbit, 2019. [Online]. Available: https://orbit.dtu.dk/en/publications/s%C3%A5dan-skaber-danmark-gr%C3%B8n-infrastruktur-til-%C3%A9n-million-elbiler-an
3. DTU and Dansk Elbil Alliance, "Smart fra start," https://danskemobilitet.dk/nyheder/pressemeddelelse/ny-rapport-bilister-sparer-penge-med-smarte-ladestandere, 2020. [Online]. Available: https://danskemobilitet.dk/nyheder/pressemeddelelse/ny-rapport-bilister-sparer-penge-med-smarte-ladestandere
4. J. Tornbjerg, "Norske elbiler skal styres smart for at skåne elnettet." [Online]. Available: https://danskemobilitet.dk/nyheder/norske-elbiler-skal-styres-smart-skaane-elnettet
5. F. Shen, S. Huang, Q. Wu, S. Repo, Y. Xu, and J. Østergaard, "Comprehensive Congestion Management for Distribution Networks Based on Dynamic Tariff, Reconfiguration, and Re-Profiling Product," *IEEE Transactions on Smart Grid*, vol. 10, no. 5, pp. 4795-4805, 2019, doi: 10.1109/TSG.2018.2868755.
6. N. Haque, A. Tomar, P. Nguyen, and G. Pemen, "Dynamic Tariff for Day-Ahead Congestion Management in Agent-Based LV Distribution Networks," (in English), *Energies,* vol. 13, no. 2, p. 318, 2020 2020, doi: http://dx.doi.org/10.3390/en13020318.
7. The Danish Energy Agency, "Fakta om flexafregning," 2019. [Online]. Available: https://ens.dk/sites/ens.dk/files/Stoette_vedvarende_energi/fakta_om_flexafregning-webtilg.pdf
8. Wallbox. "What is Smart Charging." https://wallbox.com/en_catalog/faqs-what-is-smart-charging (accessed October 14, 2020).
9. Virta. "What is smart charging?" https://www.virta.global/blog/what-is-smart-charging (accessed October 14, 2020).
10. H. Huachun, X. Haiping, Y. Zengquan, and Z. Yingjie, "Interactive charging strategy of electric vehicles connected in Smart Grids," in *Proceedings of The 7th International Power Electronics and Motion Control Conference*, 2-5 June 2012 2012, vol. 3, pp. 2099-2103, doi: 10.1109/IPEMC.2012.6259168. [Online]. Available: https://ieeexplore.ieee.org/document/6259168/
11. J. García-Villalobos, I. Zamora, J. I. San Martín, F. J. Asensio, and V. Aperribay, "Plug-in electric vehicles in electric distribution networks: A review of smart charging approaches,"

*Renewable and Sustainable Energy Reviews,* vol. 38, pp. 717-731, 2014/10/01/ 2014, doi: https://doi.org/10.1016/j.rser.2014.07.040.

12.  W. Tang, S. Bi, and Y. J. Zhang, "Online Charging Scheduling Algorithms of Electric Vehicles in Smart Grid: An Overview," *IEEE Communications Magazine,* vol. 54, no. 12, pp. 76-83, 2016, doi: 10.1109/MCOM.2016.1600346CM.

13.  N. I. Nimalsiri, C. P. Mediwaththe, E. L. Ratnam, M. Shaw, D. B. Smith, and S. K. Halgamuge, "A Survey of Algorithms for Distributed Charging Control of Electric Vehicles in Smart Grid," *IEEE Transactions on Intelligent Transportation Systems,* pp. 1-19, 2019, doi: 10.1109/TITS.2019.2943620.

14.  K. Christensen, B. N. Jørgensen, and Z. G. Ma, "A scoping review on electric vehicle charging strategies with a technical, social, and regulatory feasibility evaluation," *Energy Informatics,* (Under review).

15.  Z. Ma, K. Christensen, and B. N. Jorgensen, "Business ecosystem architecture development: a case study of Electric Vehicle home charging " *Energy Informatics,* vol. 4, p. 37, 24 June 2021 2021, Art no. 9 (2021), doi: https://doi.org/10.1186/s42162-021-00142-y.

16.  K. Christensen, Z. Ma, and B. N. Jørgensen, "Multi-agent Based Simulation for Investigating Electric Vehicle Adoption and Its Impacts on Electricity Distribution Grids and CO2 Emissions," in *Energy Informatics*, Cham, B. N. Jørgensen, L. C. P. da Silva, and Z. Ma, Eds., 2024// 2024: Springer Nature Switzerland, pp. 3-19.

17.  OK.       "Tarifmodel       3.0:       Tidsdifferentierede       timetariffer       bliver       standard." https://www.ok.dk/erhverv/produkter/el/tarifmodel (accessed March 31, 2023).

18.  K. Christensen, M. Zheng, and B. N. Jørgensen, "Technical, Economic, Social and Regulatory Feasibility Evaluation of Dynamic Distribution Tariff Designs," (in English), *Energies,* vol. 14, no. 10, p. 2860, 2021-05-27 2021, doi: https://doi.org/10.3390/en14102860.